\documentclass[11pt]{article}
\usepackage[utf8]{inputenc}
\usepackage{amsmath,amsfonts,amssymb}
\usepackage{amsthm}
\usepackage{latexsym}
\usepackage[noadjust]{cite}
\usepackage{graphicx}
\usepackage{xcolor}
\usepackage{mathtools}
\usepackage[margin=1in]{geometry}
\usepackage{thm-restate}
\usepackage{enumitem}
\usepackage[ruled,vlined,linesnumbered]{algorithm2e}
\usepackage{tikz}
\usepackage{subcaption}
\usepackage[framemethod=TikZ]{mdframed}

\definecolor{ForestGreen}{rgb}{0.1333,0.5451,0.1333}
\definecolor{DarkRed}{rgb}{0.8,0,0}
\definecolor{Red}{rgb}{1,0,0}
\usepackage[linktocpage=true,
pagebackref=true,colorlinks,
linkcolor=DarkRed,citecolor=ForestGreen,
bookmarks,bookmarksopen,bookmarksnumbered]
{hyperref}

\usepackage[capitalise]{cleveref}

\newtheorem{theorem}{Theorem}[section]
\newtheorem{corollary}[theorem]{Corollary}
\newtheorem{lemma}[theorem]{Lemma}

\newtheorem{claim}[theorem]{Claim}
\theoremstyle{definition}
\newtheorem{definition}[theorem]{Definition}

\newcommand{\E}{\mathbb{E}}
\newcommand{\dist}{\mathrm{dist}}

\title{A Near-Optimal Generalized Sorting Algorithm}

\date{\today}

\author{Sanjeev Khanna\thanks{Courant Institute, New York University, New York, NY, USA.  Supported in part by NSF award CCF-2625203 and AFOSR award FA9550-25-1-0107. Email: {\tt sanjeev.khanna@nyu.edu}.} \and Aaron Putterman\thanks{School of Engineering and Applied Sciences, Harvard University, Cambridge, Massachusetts, USA. Supported in part by a Jane Street Graduate Research Fellowship, the Simons Investigator Awards of Madhu Sudan and Salil Vadhan and AFOSR award FA9550-25-1-0112. Email: \texttt{aputterman@g.harvard.edu}.} \and Junkai Song\thanks{Courant Institute, New York University, New York, NY, USA. Supported in part by NSF award CCF-2625203. Email: \texttt{junkaisong@nyu.edu}.}}

\begin{document}
\maketitle

\begin{abstract}
We continue the study of the classic \emph{generalized sorting problem}, a basic variant of ordinary sorting. There are $n$ elements and an (initially hidden) total ordering $\pi$ over them. Unlike in ordinary sorting, however, only \emph{some} comparisons are allowed: those given by the edges of an undirected graph $G = (V, E)$ on the elements. How many comparisons are needed to find the total ordering?

In ordinary sorting, $\Theta(n \log n)$ comparisons are necessary and sufficient. For generalized sorting, however, the best known upper bound has been $\tilde{O}(n^{3/2})$ since the work of Huang, Kannan, and Khanna (FOCS 2011), who even asked whether the problem might in fact require $n^{1+\Omega(1)}$ comparisons. We show that it does not: we give a randomized algorithm that uses $O(n \log^4 n)$ comparisons in expectation. Thus, up to logarithmic factors, generalized sorting is no harder than ordinary sorting.\\

\end{abstract}
\pagenumbering{gobble}
\clearpage

\pagebreak

\tableofcontents

\pagebreak

\pagenumbering{arabic}

\section{Introduction}

We continue the study of the \emph{generalized sorting} problem. This problem is a basic variant of ordinary sorting: we are given a universe of elements $[n]$ with a hidden total ordering $\pi$, and we may query (i.e., compare) pairs of elements. Unlike in ordinary sorting, however, we are not allowed to compare \emph{all pairs} of elements. Instead, there is an underlying undirected graph $G = ([n], E)$ whose edges are the allowed comparisons; pairs of elements that are not joined by an edge cannot be compared.
To ensure that the instance is sortable, we assume that $(\pi(i), \pi(i+1)) \in E$ for every $i \in [n-1]$, or equivalently, that $G$ contains the Hamiltonian path given by $\pi$.

Generalized sorting is the special case of sorting with priced comparisons in which every comparison costs either $1$ or $\infty$. Charikar et al.~\cite{charikar2000query} initiated the study of query strategies for priced information and observed that sorting with priced comparisons appears highly non-trivial. Subsequent work studied sorting and selection when comparison costs are structured, e.g., monotone or metric~\cite{GK01, KK03, GK05}, or random~\cite{AKM08}. Variants of generalized sorting, however, date back several decades.
For instance, when $G = K_n$, this is the ordinary sorting problem, for which a myriad of algorithms and lower bounds establish that $\Theta(n\log n)$ comparisons are necessary and sufficient. When $G$ is instead the complete bipartite graph $K_{n/2, n/2}$, this recovers the famous ``sorting nuts and bolts'' problem, where only elements on \emph{different sides} of the bipartition can be compared (nuts cannot be compared with nuts, nor bolts with bolts). Through a sequence of works from the 1990s \cite{matching1, alon1996matching, komlos1998matching}, a tight deterministic bound of $\Theta(n \log n)$ queries was ultimately established for this problem as well.\footnote{An $O(n \log n)$ randomized algorithm is easy to obtain for this setting via a QuickSort style partitioning scheme.}
These works rely crucially on the \emph{complete bipartite structure} of the comparison graph.

In many natural settings, however, the comparison graph has no such structure. As a motivating example from \cite{huang2011algorithms}, consider several candidates applying for the same job, together with a set of references for these candidates. Two candidates are \emph{comparable} if some reference knows both of them. Clearly then, there is no guarantee that \emph{every pair} of candidates is comparable.

As observed by Huang, Kannan, and Khanna \cite{huang2011algorithms}, the standard approaches to sorting break down when only some comparisons are allowed. For instance, a QuickSort-style algorithm needs a pivot that can be compared with most other elements, but an arbitrary graph may have no such pivot. Similarly, a MergeSort-style algorithm splits the elements into subproblems and sorts each of them recursively. This also breaks down in our setting for a subtle reason, which is that \emph{a subproblem may not be sortable on its own}. Indeed, when we arbitrarily restrict to smaller subproblems, elements that are consecutive \emph{within a subproblem} need not be comparable. 

More generally, algorithms for ordinary sorting implicitly rely on the fact that at every stage, some comparison is \emph{balanced}: whichever way it is answered, it eliminates a constant fraction of the orderings that are still consistent with the answers so far. Making such a comparison at every step sorts the elements using $O(\log n!) = O(n \log n)$ comparisons. In contrast, when comparisons are restricted, there may be intermediate stages at which no allowed comparison is balanced, which makes generalized sorting considerably more challenging.

As a result, progress on generalized sorting for \emph{arbitrary graphs} has been slow.
 For a decade after \cite{charikar2000query}, no $o(n^2)$-query algorithm was known, until Huang, Kannan, and Khanna \cite{huang2011algorithms} showed that generalized sorting is possible with $\tilde O({n^{3/2}})$ queries. Subsequent works extended \cite{huang2011algorithms} in several directions, but none improved the $\tilde O({n^{3/2}})$ bound for arbitrary graphs, which has stood since 2011. 
For instance, Kuszmaul and Narayanan \cite{kuszmaul2022stochastic} showed that when $G$ has $m$ edges, generalized sorting can be done with $\widetilde{O}(\sqrt{mn})$ queries; however, when $m = \Omega(n^2)$, this provides \emph{no} improvement over \cite{huang2011algorithms}. They also showed that when $G$ is an \emph{Erd\"os-Renyi} random graph (with the promised Hamiltonian path added), generalized sorting can be done with $\tilde O(n)$ queries in expectation.

In a different direction, Lu, Ren, Sun, and Zhang \cite{lu2020generalized} studied \emph{generalized sorting with predictions}, where one is given a \emph{prediction} function $P: E \rightarrow \{ \leftarrow, \rightarrow\}$ that predicts the orientation of every edge of $G$. When $P$ makes only $w$ errors, \cite{lu2020generalized} shows that generalized sorting can be done in $O(n\log n + w)$ queries. The difficulty of the general problem has also motivated the study of special cases of generalized sorting on structured graphs \cite{banerjee2016sorting, biswas2017improved, chan2023generalized, jiang2023algorithms, goswami2022instance, goswami2023algorithm}, but an $n^{3/2 - \Omega(1)}$-query algorithm for general graphs has remained elusive.

Taken together, these results leave open the central question in the study of generalized sorting. The only known lower bound is the same $\Omega(n \log n)$ bound from ordinary sorting. In fact, motivated by the lack of progress, Huang, Kannan, and Khanna \cite{huang2011algorithms} even asked whether it can be strengthened to $\Omega(n^{1+\varepsilon})$ for some constant $\varepsilon > 0$. In this work, we address the following question:
 
\begin{center}
    \emph{Is generalized sorting on arbitrary graphs polynomially harder than ordinary sorting? Can one perform generalized sorting in $n^{3/2 - \Omega(1)}$ queries, or even in $\tilde{O}(n)$ queries?}
\end{center}

\subsection{Our Main Result}

Our main result nearly settles this question: we show that the query complexity of generalized sorting matches that of ordinary sorting up to polylogarithmic factors. Our main theorem is the following:
 
\begin{theorem} \label{thm:main}
    There is a randomized algorithm that, for every graph $G = (V, E)$ on $n$ vertices and every hidden ordering $\pi$ of $V$ such that $(\pi(i), \pi(i+1)) \in E$ for all $i \in [n-1]$, recovers $\pi$ using $O(n \log^{4} n)$ queries in expectation.
\end{theorem}
 
In particular, Theorem~\ref{thm:main} answers the question of \cite{huang2011algorithms}: generalized sorting is \emph{not} polynomially harder than ordinary sorting.  
Perhaps surprisingly, despite the long history of the problem, our algorithm turns out to be relatively simple.

\paragraph{A Brief Overview of Techniques}
Our starting point is the prediction-repair algorithm of \cite{lu2020generalized}. Indeed, as mentioned above, \cite{lu2020generalized} shows that, given a prediction function (which predicts an orientation for every edge in $E$) that makes only $w$ errors, one can perform generalized sorting in only $O(n\log(n) + w)$ many comparisons. This already suggests a natural starting point: can one show that in only $\tilde{O}(n)$ queries, one can construct a predictor which makes only $\tilde{O}(n)$ errors?

Our main result is an algorithm which exactly achieves this above goal, though the actual construction of the predictor may be slightly counterintuitive. To describe this, we begin with the key lemma which underlies the construction of this predictor:

\begin{lemma}
    For any subset of $m$ edges $F \subseteq E$, there is a randomized procedure that partitions $F$ into a ``removed'' set $T$ and the edges $F \setminus T$, along with predictions for these latter edges, such that:
    \begin{enumerate}
        \item $|T| \leq m/2$.
        \item The expected number of incorrect predictions on $F \setminus T$ is $O(n)$.
        \item The total number of queries made is $O(n \log^3(n))$.
    \end{enumerate}
    Moreover, the procedure \emph{does not} require $(V, F)$ to contain a Hamiltonian path. 
\end{lemma}

It is worth making several remarks on the above lemma. First, it works for an \emph{arbitrary subset} of the original edges in the graph $G$, \emph{even if} the chosen set of edges does not contain a Hamiltonian path. Beyond this, it produces a good prediction of edge orientations (with only $O(n)$ expected errors) for \emph{most} of the edges in the set $F$. For some edges (those in the ``removed set'' $T$), it will neglect to produce a prediction, and defers the prediction for such edges until future rounds. 

With the above lemma in hand, it is not hard to see how we produce the final predictor. We start with the entire edge set, i.e., $F = E$, and apply the above lemma. This makes $O(n \log^3(n))$ queries, and produces a prediction over at least half of the edges in $E$ which makes only $O(n)$ errors. The set $T$ still has no predictions made, so we then repeat the above lemma with the choice $F = T$, and so on.

The key subtlety emerges in proving this lemma, which follows a type of ``prediction refinement'' procedure. In more detail, given the edge set $F$, we start by sampling a ``coarse prediction'' of the orientations of edges in $F$. This coarse prediction is not very accurate and makes many errors, say, on the order of $\approx \frac{m}{\log^2(n)}$. There are then two ways that one could hope to decrease the number of errors:
\begin{enumerate}
    \item We can try to \emph{update} the prediction to be better on its mispredicted edges.
    \item Or, we can try to locate regions of the graph where the mispredictions are \emph{denser than average}, and remove these.
\end{enumerate}
Our approach only tries to implement the second objective above, repeatedly removing regions with dense errors, and in so doing, driving down the error rate in the remaining edges.
This takes the form of an iterative algorithm where in the $i$th round, we identify a batch of roughly $\approx \frac{m}{2\log(n)}$ many edges, such that, after removing these edges, the number of remaining mispredictions decreases by a factor of $1/2$. One can then see that across $\log(n)$ rounds of this procedure, the number of mispredictions decreases to $\approx O(n)$, while the number of removed edges scales as $\approx \frac{m}{2 \log(n)} \cdot \log(n) = m/2$.

Of course, this still leaves the task of designing this single round procedure, which we dub the \emph{error isolation protocol}. To simplify the exposition here, we suppose that we are working with a \emph{bipartite graph}, where the prediction guesses that all edges go from $L$ to $R$.
Roughly speaking, this protocol then operates by \emph{randomly sampling} a set of mispredicted edges (edges that point from right to left), and storing the right hand vertices of every sampled edge in a set $S$. It then queries all of the edges to the \emph{left} neighbors of these vertices in $S$. Ultimately, the procedure must produce a collection of edges which has disproportionately many errors, and it does this in two ways:
\begin{enumerate}
    \item When querying the left neighbors of each $s \in S$, if an edge $(\ell, s)$ is mispredicted, we remove \emph{all} of $\ell$'s edges. 
    \item Then, we choose a permutation $\tau$ over the vertices $S$. For each $\ell \in L$, we compute its score to be the smallest rank (in $\tau$) of any of $\ell$'s neighbors in $S$ (if there are no neighbors, we give it rank $|S|+1$). For every $r \in R$, we then remove all edges to $r$'s highest scoring neighbors.
\end{enumerate}
The first condition above has a very intuitive interpretation, as it simply captures that a left vertex may have a ``high density'' of errors among its sampled incident edges. The second condition is less intuitive; but is roughly designed with the goal of capturing errors that are \emph{not observed} in our random sampling. This is because if a left vertex has a high score, then none of the sampled vertices appearing \emph{early} in $\tau$ are neighbors of this left vertex. It turns out that this provides a valuable signal that we should suspect that our random sampling is not as accurate for this vertex.

Surprisingly, the heart of the analysis in \cref{sec:errorIsolation} shows that there \emph{exists} a choice of this permutation $\tau$ over $S$ for which the edge removal procedure above succeeds in reducing the number of remaining errors by a factor of $2$.
Of course, there still remain many details in the procedure, including: (a) how do we find this optimal choice of $\tau$? (b) how do we ensure that not too many queries are made? (c) how do we locate the mispredicted edges? and so on. We defer these complete descriptions to the (brief) technical sections below, but nevertheless, this discussion above captures the overarching structure of our algorithm.

\subsection{Future Directions}

We highlight three natural directions for future work.

\begin{enumerate}
    \item \textbf{Closing the remaining gap.} Our algorithm uses $O(n \log^4 n)$ queries, an $O(\log^3 n)$ factor more than the $\Omega(n \log n)$ lower bound for ordinary sorting. Can generalized sorting be done with $O(n \log n)$ queries on every graph, or do some graphs require $\omega(n \log n)$ queries?

    \item \textbf{Deterministic algorithms.} Similar to previous algorithms for sorting on arbitrary graphs \cite{huang2011algorithms, kuszmaul2022stochastic}, our algorithm is randomized. What is the deterministic query complexity of generalized sorting? In particular, is there a deterministic algorithm that uses $\tilde O(n)$ queries? Even for nuts and bolts, where a simple randomized variant of QuickSort uses $O(n \log n)$ comparisons in expectation, matching this bound deterministically took a sequence of works \cite{matching1, alon1996matching, komlos1998matching}.

    \item \textbf{Polynomial-time algorithms.} We focus on query complexity, and as stated, our algorithm does not run in polynomial time. Is there a polynomial-time algorithm that uses $\tilde O(n)$ queries, or even $n^{3/2-\Omega(1)}$ queries? For comparison, the $\tilde{O}(n^{3/2})$-query algorithm of \cite{huang2011algorithms} runs in polynomial time.
\end{enumerate}

\section{Overview}\label{sec:prelim}
We begin by defining the distance between two orderings or predictions.
\begin{definition}
    For a graph $G = (V, E)$, two total orderings $\pi, \sigma$ over $V$, and a subset of edges $F\subseteq E$, we define 
    \[
    \mathrm{dist}_F(\pi, \sigma) = \sum_{(u,v) \in F} \mathbf{1}[\mathbf{1}[u\prec_{\pi} v] \neq \mathbf{1}[u\prec_{\sigma} v]].
    \]
    Likewise, we extend this definition to \emph{prediction} functions, which map every edge $(u,v)\in E$ to an orientation $u\to v$ or $v\to u$. For such a function $P$, we define 
    \[
    \mathrm{dist}_F(\pi, P) = \sum_{e=(u,v) \in F} \mathbf{1}[\mathbf{1}[u\prec_{\pi} v] \neq \mathbf{1}[P(e)=u\to v]].
    \]
    We write $\dist_G(\pi,\sigma)=\dist_E(\pi,\sigma)$ and $\dist_G(\pi,P)=\dist_E(\pi,P)$.
\end{definition}

Our algorithm has two main components. The first, which is our main contribution, constructs a prediction function with $\tilde O(n)$ expected errors using $\tilde O(n)$ queries. The second, due to \cite{lu2020generalized}, recovers the hidden ordering $\pi$ from a prediction function with $w$ errors using $\tilde O(n+w)$ queries. We describe these components below.

\paragraph{Prediction construction.} Our algorithm proceeds in phases. In each phase, it fixes predictions on a subset of edges and postpones the remaining edges to subsequent phases. The progress made in each phase is summarized by the following lemma.

\begin{lemma} \label{lem:phase}
Let $G=(V,E)$ be an undirected graph with a hidden total ordering $\pi$ over its vertices. For any subset of edges $F$, there is a randomized algorithm that makes $O(n\log ^3n)$ comparisons and partitions $F$ into a postponed set $F'$ and a set of predicted edges $F\setminus F'$ such that $|F'|\leq |F|/2$ and the expected number of incorrect predictions on $F\setminus F'$ is at most $O(n)$.
\end{lemma}

By repeatedly applying the lemma to the postponed edges, the number of unresolved edges decreases by half in each phase. Therefore, after $O(\log n)$ phases, all edges in $G$ are predicted. Combining these predictions yields a prediction function $P$ with $O(n\log n)$ many expected errors, using $O(n\log^4 n)$ comparisons in total.

\paragraph{Prediction repair.}

After obtaining a prediction function with few errors, we apply the \emph{prediction repair} algorithm due to \cite{lu2020generalized}. We remark that this is the only step that requires the graph to contain the Hamiltonian path induced by the hidden total order $\pi$.

\begin{lemma}[\cite{lu2020generalized}] \label{lem:repair}
    Let $G$ be an undirected graph on $n$ vertices and $m$ edges with a hidden total ordering $\pi$ over its vertices such that all consecutive vertices in $\pi$ are adjacent in $G$. Suppose there is a prediction function $P$ over the edges of $G$ such that 
    \[
    \mathrm{dist}_{G}(\pi, P) \leq w.
    \]
    Then there is a randomized algorithm that, given $G$ and $P$, makes at most $O(n\log n + w)$ queries in expectation and recovers the true ordering $\pi$.
\end{lemma}

Combining the prediction construction and repair algorithms proves \cref{thm:main}. The remainder of the paper is devoted to proving \cref{lem:phase}.

\section{Repeated Error Isolation for Near-Optimal Generalized Sorting}

In this section, we describe our prediction construction algorithm in detail. The key idea underlying our algorithm is to iteratively \emph{refine} a prediction of edge orientations by gradually removing sets of edges that are ``denser'' with errors than the remaining edges. Throughout this section, we suppose that we are working with a set $F$ of $|F|=f$ edges from the original generalized sorting instance. We begin by describing a procedure that produces a \emph{coarse prediction} of the orientations of the edges in $F$.

\subsection{Building a Coarse Predictor}

In this section, we describe the \emph{coarse prediction} routine, which uses $\tilde O(n)$ queries to construct a prediction for $F$ with $O(f/\log^2 n)$ errors. If $f \leq 2^{24}n\log^2 n$, we simply query all edges in $F$ and terminate. Otherwise, we sample $2^{32}  n \log^3n$ edges uniformly at random from $F$ (with replacement), and then let $\sigma$ be \emph{any} permutation of the vertices $V$ which agrees with the reported orientation of \emph{every} sampled edge. Note that such a permutation must exist, since the true permutation $\pi$ agrees with all reported orientations. For this permutation $\sigma$, we have the following claim:

\begin{claim}\label{clm:goodCoarsePredictor}
    With probability $1 - 2^{-10n}$ over the random sample, it is the case that 
    \[
    \mathrm{dist}_{F}(\pi,\sigma) \leq \frac{f}{2^{28} \log^2 n}.
    \]
\end{claim}

\begin{proof}
    Fix \emph{any} permutation $\sigma'$ with $\mathrm{dist}_{F}(\pi,\sigma') > \frac{f}{2^{28}\log^2 n}$. A uniformly random edge of $F$ that we sample has probability at least $\frac{1}{2^{28} \log^2(n)}$ of disagreeing with orientation $\pi$. Thus, the probability that $\sigma'$ remains consistent with $2^{32}  n \log^3 n$ sampled edges (and therefore could be returned as the coarse predictor) is at most 
    \[
    \left (1 - \frac{1}{2^{28} \log^2n} \right )^{2^{32} n \log^3n} \leq e^{-16 n \log n}.
    \]
    Taking a union bound over all $n!$ possible choices of $\sigma'$, we obtain
    \[
    \Pr\left [\mathrm{dist}_{F}(\pi,\sigma) > \frac{f}{2^{28} \log^2(n)} \right] \leq n! \cdot e^{-16 n \log(n)} \leq 2^{-10n}.
    \]
    which yields the desired claim.
\end{proof}

For the rest of this section, we  condition on the good event that $\mathrm{dist}_{F}(\pi,\sigma) \leq \frac{f}{2^{28} \log^2(n)}$.

\subsection{Building a Bounded-Degree Bipartite Prediction Graph}

We next transform $G[F]$ into a bipartite graph with maximum degree $O(f/n)$, which will be useful in our algorithm and analysis.

To start, we duplicate the vertex set $V$, creating a left copy $L$ and a right copy $R$. For every edge $(u,v) \in F$ with $u\prec_{\sigma} v$, we add an edge between the copy of $u$ in $L$ and the copy of $v$ in $R$. Thus, $\sigma$ predicts every edge to be oriented from $L$ to $R$.

Next, let $d = \lceil f / n \rceil$ denote the \emph{average degree} of the bipartite graph. If any left or right vertex has more than $d$ incident edges, we split the edges into groups of size at most $d$, making a separate copy of the vertex for each group. We denote the final resulting bipartite graph from this procedure by $H=(L\cup R,F)$. We have the following claims about this graph:

\begin{claim}\label{clm:Hproperties}
    Let $H$ be constructed as above. Then, all of the following hold:
    \begin{enumerate}
        \item Every vertex in $H$ has degree at most $d$.
        \item There are at most $2n$ left vertices and $2n$ right vertices.
        \item $n \geq d > 2^{24} \log^2(n)$ and $f \leq nd \leq 2f$.
    \end{enumerate}
\end{claim}

\begin{proof}
The first item is immediate. To see the second item, observe that the number of left vertices is bounded by $n + \frac{f}{d} \leq 2n$, and similarly for right vertices.  For the third item, it follows from our assumption that $f > 2^{24} n \log^2(n)$ and the definition of $d$.
\end{proof}

\subsection{Removing Right Vertices with Too Many Errors}
In the bipartite graph $H=(L\cup R,F)$, we say that an edge $(u,v)$ with $u\in L,v\in R$ is a \emph{bad edge} if if its orientation is incorrectly predicted by $\sigma$, i.e., $u\succ_\pi v$. Let $w=\dist_F (\pi,\sigma)$ be the total number of bad edges. The next step of our algorithm identifies and removes right vertices that are incident to too many bad edges.

To do this, we take a new uniformly random sample of $2^{32} n \log^3 n$ many edges from $F$, and query their true orientations. We \textbf{flag} a vertex $v \in R$ if at least $2^9 \log n$ sampled edges incident to $v$ are bad. We have the following claim about flagged vertices:

\begin{claim}\label{clm:flaggedVerticesBadEdges}
    With probability $1 - n^{-10}$, the following hold:
    \begin{enumerate}
        \item Every flagged vertex $v \in R$ is incident to at least $\frac{d}{2^{24}\log^2 n}$ bad edges. 
        \item Every unflagged vertex $v \in R$ is incident to most $\frac{4d}{2^{24}\log^2 n}$ bad edges.
    \end{enumerate}
\end{claim}

\begin{proof}
    Fix a right vertex $v \in R$ with $b_v$ many bad incident edges. The number of bad edges incident to $v$ in the sample is then exactly distributed as a binomial random variable with mean
    \[
    \mu=2^{32} n \log^3n \cdot \frac{b_v}{f}.
    \]
    If $b_v < \frac{d}{2^{24}\log^2 n}$, then $\mu\leq 2^{8} \log n$. If $b_v > \frac{4d}{2^{24}\log^2 n}$, then $\mu\geq 2^{10} \log n$. Since the threshold is $2^9\log n$, by standard Chernoff bounds, vertices in the first case are unflagged with probability $1 - 1/n^{12}$, and vertices in the second case are flagged with probability $1 - 1 / n^{12}$. By taking a union bound over all vertices, we then obtain the stated claim. 
 \end{proof}

We now \emph{remove} all edges incident to flagged vertices from $H$. The following claim bounds the number of removed edges.

\begin{claim} \label{clm:remove}
    Assuming the good events in \cref{clm:goodCoarsePredictor} and \cref{clm:flaggedVerticesBadEdges}, the number of removed edges is at most $f/16$.
\end{claim}

\begin{proof}
    By \cref{clm:goodCoarsePredictor}, we know that the \emph{total} number of bad edges is bounded by $\frac{f}{2^{28} \log^2(n)}$. At the same time, by \cref{clm:flaggedVerticesBadEdges}, every flagged vertex has at least $\frac{d}{2^{24} \log^2(n)}$ many distinct bad edges incident to it. Thus, the number of flagged vertices is at most
    \[
    \frac{f/(2^{28}\log^2 n)} {d/(2^{24}\log^2 n)} =\frac{f}{16d}.
    \]
    As every vertex has degree at most $d$, removing all edges incident to flagged vertices deletes at most $f/16$ edges.
\end{proof}

\subsection{The Error Isolation Protocol}\label{sec:errorIsolation}

In this section, we describe the \emph{error isolation} protocol. We use positive integers $s$ and $t$ as parameters of our protocol, and use $w$ to denote the number of bad edges (errors) in the current graph $H$ under consideration. The steps of the protocol are:
\begin{enumerate}
    \item We first sample $s$ uniformly random \emph{bad edges} (with replacement), by repeatedly querying uniformly random edges of $H$ until a bad edge is found. For each bad edge, we store its right endpoint in a (multi)set $S$. Thus, at the termination of this step, $|S| = s$.
    \item For every vertex $v \in S$, we query all edges of $H$ incident to $v$. For any left vertex $u \in L$ where we discover an incorrectly oriented edge $(u,v)$ (that is, $v\prec_{\pi} u$), we say that $u$ is a \emph{marked} vertex.
    \item Fix an arbitrary permutation $\tau$ over the sampled right vertices $S$. For each left vertex $u \in L$, we define its score under $\tau$ as the \emph{minimum} rank in $\tau$ over all of its neighbors in $S$. I.e., 
    \[
    \mathrm{Score}_{\tau}(u) = \min_{v \in \Gamma(u) \cap S} \tau(v).
    \]
    If $u$ has no neighbors in $S$, we simply define its score to be $s+1$.
    \item For every right vertex $v \in R$, we select its $t$ \emph{highest score} left neighbors, breaking ties arbitrarily. Let $Q_v$ denote the set of edges connecting $v$ to its selected neighbors.
    \item Finally, let $C_{\tau}$ contain (1) all edges incident to a marked left vertex $u \in L$, and (2) for every $v \in R$, the $t$ edges going to its highest score left neighbors. I.e.,
    \[
    C_{\tau} = \bigcup_{\text{marked }u \in L} E_H[u, \Gamma(u)] \cup \bigcup_{v \in R} Q_v. 
    \]
\end{enumerate}

As we will see in the following discussion, an appropriate choice of $\tau$ ensures that removing $C_\tau$ \emph{significantly} reduces the number of bad edges. The key claim that we show is the following. Crucially, choosing $s, t$ such that $(s+1) t \geq 32|L|$ ensures that removing $C_{\tau^*}$ reduces the number of remaining errors by a large constant factor.

\begin{lemma}\label{lem:remainingErrorBound}
For each sampled multiset $S$, let $\tau^*$ denote the ordering of vertices in $S$ induced by their relative order in the true permutation $\pi$. Then, we have
\[
\E_S[\text{number of bad edges in }(H \setminus C_{\tau^*})] \leq  \frac{|L|\cdot w}{(s+1)t}.
\]
\end{lemma}

\begin{proof}
We say a left vertex $u$ is \emph{late} with respect to a right vertex $v$ if every sampled neighbor of $u$ comes after $v$ in the true permutation $\pi$. For a right vertex $v$, let $b_v$ denote the number of bad edges incident to $v$, and let $\ell_v$ denote the number of its late neighbors.

First, observe that every bad edge $(u,v)$ with $u$ unmarked must have $u$ late with respect to $v$. Indeed, since $u$ is unmarked, all of its sampled neighbors in $S$ come after $u$ in $\pi$, while the bad edge $(u,v)$ implies that $v\prec_{\pi} u$. 

Moreover, under $\tau^*$, every late neighbor of $v$ has strictly higher score than every non-late neighbor. 
To see this, for any right vertex $v$, define
\[
c_v=|\{i\in[s]:x_i\preceq_\pi v\}|,
\]
where $x_1, \dots x_s$ are the vertices in $S$. Thus, $c_v$ counts the number of sampled vertices coming \emph{at or before} $v$ in the true order. Crucially, if $u$ is late with respect to $v$, all sampled neighbors come after $v$, and so $\mathrm{Score}_{\tau^*}(u) > c_v$. Otherwise, if $u$ is \emph{not late} with respect to $v$, some sampled neighbor of $u$ is at or before $v$, and therefore $\mathrm{Score}_{\tau^*}(u) \leq c_v$, yielding the claim.

With these claims, we can see that if $\ell_v \leq t$, then \emph{all} bad edges $(u,v)$ with $u$ unmarked are removed in the set of the $t$ highest score vertices. This is because every such neighbor $u$ is late, and every late neighbor has a higher score than any non-late neighbor. Since there are at most $t$ late neighbors, all of them are among the $t$ with the highest scores. Therefore, after removing the $t$ highest score neighbors, the number of bad edges $(u,v)$ incident to $v$ with $u$ unmarked is at most
\[
b_v \cdot \mathbf{1}[\ell_v > t] \leq \frac{b_v \ell_v}{t}.
\]
In particular, we can then bound the \emph{total number} of such bad edges as 
\[
\sum_{v \in R}\frac{b_v \ell_v}{t} = \frac{w}{t} \sum_{v \in R} \frac{b_v \ell_v}{w} = \frac{w}{t} \E[\ell_X],
\]
where $X$ is a random right vertex drawn from the same distribution as each sampled element of $S$, namely $\Pr[X=v]=b_v/w$.

Finally, it remains to bound $\E[\ell_X]$. Fix any left vertex $u \in L$, and consider the $s+1$ independent samples consisting of $X$ and the vertices in $S$. The event that $u$ is a late neighbor of $X$ requires $X$ to come before every sampled neighbor of $u$ in $\pi$. Equivalently, among the samples that are neighbors of $u$, $X$ must be the (unique) smallest in $\pi$. As $X$ and the samples in $S$ come from the same distribution, this event has probability at most $\frac{1}{s+1}$. Summing over all left vertices, we obtain
\[
\E[\ell_X] \leq \frac{|L|}{s+1}.
\]
Plugging this into our earlier bound, we have that the expected (over our choice of $S$) number of remaining bad edges  $(u,v)$ with $u$ unmarked is at most
\[
\frac{|L|}{s+1} \cdot \frac{w}{t}.
\]
Equivalently,
\[
\E_S[\text{number of bad edges in }(H \setminus C_{\tau^*})] \leq \frac{|L|\cdot w}{(s+1)t}.
\]
\end{proof}

\subsection{Invoking the Error Isolation Protocol}

We now invoke the error isolation protocol. Throughout this section, we use $H$ to denote the \emph{current} graph, consisting of edges that have not yet been removed in \cref{lem:remainingErrorBound}.

We fix the parameters of the error isolation protocol as
\[
s := \frac{2^{18} \cdot n \log n}{d}, \quad \quad t := \frac{d}{2^{12} \log n}.
\]
Importantly, we see that $st = 64n > 32 \cdot |L|$ which will allow us to conclude a significant reduction in the number of errors when applying \cref{lem:remainingErrorBound}. With these definitions, we are now ready to define a \emph{single pass} of our error isolation procedure.

\paragraph{A Single Error Isolation Pass.}

We outline the procedure below. This procedure starts by sampling random edges, hoping to find enough bad edges to run the algorithm from \cref{sec:errorIsolation} If not enough bad edges are found, we are already in a good case. Otherwise, if it finds enough bad edges, it carries out \cref{sec:errorIsolation}'s protocol. It remains still to identify the permutation $\tau^*$ from \cref{lem:remainingErrorBound}. This is done by using an \emph{additional} sample of bad edges to effectively ``certify'' whether any permutation behaves similarly to $\tau^*$.

\begin{enumerate}
    \item To start, we repeatedly sample \emph{uniformly random} edges from $H$ and query their true orientations. The first $s$ bad edges found are called the \emph{training edges}, and the subsequent $2^{10}  s \log n$ bad edges are called the \emph{validation edges}. Note that if we fail to obtain $s+2^{10}  s \log n$ bad edges within $16d \cdot (s + 2^{10}s \log n)$ many samples and queries, then we terminate prematurely. In the following steps, we assume the sampling succeeds.
    \item Next, we perform the algorithm in \cref{sec:errorIsolation} using the first $s$ bad edges (the training edges). Let $S$ be the multiset of right endpoints. We query all edges incident to vertices in $S$, marking every left vertex incident to a queried bad edge. Then, for every ordering $\tau$ of $S$, we construct the candidate removal set $C_\tau$.
    \item Finally, it remains to \emph{select} one of the candidate removal sets. For this, we use the sample of $2^{10} \cdot s \log n$ bad, validation edges. We choose the candidate removal set $C_{\tau'}$ that misses the \emph{fewest} validation edges.
\end{enumerate}

This completes the description of a single pass. We now analyze its properties and show that it successfully removes a large number of bad edges.

\paragraph{Analyzing a Single Pass.}

We first show that if at least $n$ bad edges remain in $H$, then the sampling step succeeds with constant probability.

\begin{claim}\label{clm:notEnoughBad}
    Suppose the current graph $H$ contains $w\geq n$ bad edges. Then, with probability at least $15/16$, among the sample of $16d \cdot (s + 2^{10}s \log n)$ edges, there will be $s+2^{10} s \log n$ bad edges.
\end{claim}

\begin{proof}
    Since $H$ contains at most $nd$ edges and at least $n$ bad edges, the expected number of samples needed to collect $s+ 2^{10} \cdot s \log n$ bad edges is at most $d \cdot (s+2^{10} s \log n)$ many edges. By a simple Markov bound, the probability that this takes more than $16d\cdot(s+2^{10}s\log n)$ queries is at most $1/16$, proving the claim.
\end{proof}

Next, we can observe that the bad edges collected by the sampling step are independent and uniformly distributed among the bad edges, and therefore is in line with the sampling procedure in \cref{sec:errorIsolation}.  In particular, this allows us to immediately conclude the following:
\begin{corollary}\label{cor:optCandidate}
   Let $H$ be the current bipartite graph, and let $\tau^*$ denote the true relative order of the sampled multiset $S$. Then,
   \[
   \E_S[\text{number of bad edges in }(H \setminus C_{\tau^*})] \leq  \frac{|L|\cdot w}{(s+1)t} \leq \frac{w}{32}.
   \]
\end{corollary}
Note that here we have used the fact that $s \cdot t \geq 64n$, and that the number of left and right vertices in $H$ is at most $2n$.

However, the algorithm does not know \emph{true} permutation $\tau^*$, but instead choose the permutation $\tau'$ for which $C_{\tau'}$ misses the fewest validation edges. For this $\tau'$, we have the following claim:

\begin{claim}\label{clm:decreaseInBad}
    Let $\tau'$ denote the permutation  whose candidate removal set $C_{\tau'}$ misses the fewest of the $2^{10} s \log(n)$ validation edges. If the current graph $H$ contains $w\geq n$ bad edges, then after a single pass,
    \[
    \E[\text{number of remaining bad edges in }H] \leq \frac{w}{4}.
    \]
\end{claim}

\begin{proof}
    By \cref{cor:optCandidate}, we know that there exists a permutation $\tau^*$, for which $C_{\tau^*}$ misses a $\leq 1 / 32$ fraction of the validation edges in expectation.

    Now, let us fix any candidate permutation $\tau$. The fraction of the $2^{10} s \log n$ validation edges that are missed by $C_{\tau}$ is, in expectation, equal to the \emph{true fraction} of bad edges missed by $C_{\tau}$. We call this value $\rho_{\tau}$ the \emph{missing probability}, and denote its estimate from the sampled validation edges by $\hat{\rho}_{\tau}$. By applying a standard Hoeffding bound,
    \[
    \Pr[|\rho_{\tau} - \hat{\rho}_{\tau}| > 1/16] \leq 2e^{-8s \log n}.
    \]
    Taking a union bound over all $s!$ candidate permutations, we obtain that the probability that $|\rho_{\tau} - \hat{\rho}_{\tau}| > 1/16$ for any $\tau$ is bounded by $s! \cdot 2e^{-8s \log n} \leq 1/32$.

    Now, conditioned on $|\rho_{\tau} - \hat{\rho}_{\tau}| \leq 1/16$ for every $\tau$, we can see that for the selected permutation $\tau'$, we have
    \[
    \rho_{\tau'} \leq \rho_{\tau^*} + 2 \cdot 1/16.
    \]
    Note that if the bad case of \cref{clm:notEnoughBad} occurs, or the bad case above, where $|\rho_{\tau} - \hat{\rho}_{\tau}| > 1/16$ for some $\tau$, we can bound the expected fraction of missed bad edges by $\tau'$ by $1$.

    Thus, in total, we can see that 
    \[
    \E[\rho_{\tau'}] \leq \underbrace{\frac{1}{32}}_{\text{true candidate }\tau^*} + \underbrace{\frac{1}{8}}_{\text{error bound on }\hat{\rho}} + \underbrace{\frac{1}{32}}_{\text{prob. of failing to estimate }\hat{\rho}} + \underbrace{\frac{1}{16}}_{\text{prob. of not enough bad edges}}  \leq \frac{1}{4}.
    \]
\end{proof}

We can now apply this repeatedly to reduce the number of bad edges to $O(n)$.

\begin{lemma}\label{lem:repeatedIsolationPass}
    Let $H$ be the current graph with at most $n^2$ edges, and let $\sigma$ be the coarse prediction. We apply the single error isolation pass $\log n$ times, removing the selected edge removal set $C_{\tau'}$ from $H$ in each iteration. After these $\log n$ iterations, 
    \[
   \E[\text{number of remaining bad edges in }H]  = O(n).
    \]
\end{lemma}

\begin{proof}
    Indeed, by \cref{clm:decreaseInBad}, each iteration of the above procedure reduces the expected number of bad edges from $w$ to at most $w/4$ in expectation, provided that $w\geq n$. Thus, we can unconditionally bound the expected number of remaining bad edges by $\frac{w}{4} + n$.

    In particular, after $\log n$ iterations, the expected number of remaining bad edges in $H$ is at most
    \[
    \frac{|E(H)|}{4^{\log n}} + n \cdot \left ( 1 + \frac{1}{4} + \frac{1}{16} + \dots  \right ) \leq \frac{n^2}{4^{\log n}} + n \cdot \frac{4}{3} = O(n).
    \]
\end{proof}

This is extremely valuable; after $\log n$ iterations, we have essentially ``purified'' our graph $H$, getting rid of most of the bad edges. However, it remains to show that in the process of removing these candidate removal sets $C_{\tau'}$, we have \emph{not removed too many edges from }$H$. Indeed, a trivial way to purify $H$ would simply have been to remove \emph{all} of its constituent edges. In the following, we bound the total number of edges removed by our algorithm.

\paragraph{Bounding the Number of Removed Edges.}

In this section, we will show the following:

\begin{claim}\label{clm:edgeRemovalBound}
    Let $H$ initially contain $f$ edges. Then, with probability $1 - 2 / n^{10}$, the flagging step and the subsequent $\log n$ error isolation passes together remove at most $f/2$ edges. 
\end{claim}

\begin{proof}
We condition on the good events of \cref{clm:goodCoarsePredictor} and \cref{clm:flaggedVerticesBadEdges}. By \cref{clm:flaggedVerticesBadEdges}, every right vertex in $H$ always has at most $\frac{4d}{2^{24}\log^2 n}$ incident bad edges. Thus, in a single iteration of the error isolation protocol, the $s$ sampled right vertices mark at most 
\[
s \cdot \frac{4d}{2^{24}\log^2 n}
\]
many left vertices. In turn, each of these left vertices ends up removing their entire neighborhoods, leading to a total of 
\[
d \cdot s \cdot \frac{4d}{2^{24}\log^2 n}
\]
many removed edges. The remaining edges in each candidate removal set come from the $t$ highest scoring vertices incident to each right vertex. Because there are at most $2n$ right vertices, and $t = \frac{d}{2^{12}\log n}$, this constitutes at most $\frac{nd}{2^{11}\log n}$ many removed edges. Therefore, in each of the $\log n$ rounds, the total number of removed edges is at most
\[
d \cdot s \cdot \frac{4d}{2^{24}\log^2 n} + \frac{nd}{2^{11}\log n} \leq 2^{18} n \log n \cdot \frac{4d}{2^{24}\log^2 n} + \frac{f}{2^{10}\log n} \leq \frac{f}{4 \log n},
\]
where we have used that $s \cdot d = 2^{18} n\log n$, and that $nd \leq 2f$. Summing across all $\log n$ rounds, this implies that at most $f/4$ edges are removed by the error isolation procedure. 

Finally, by including the $\leq f/16$ edges removed in \cref{clm:remove}, we obtain the stated bound.
\end{proof}

\paragraph{Bounding the Number of Queries Performed.}

The final piece of our analysis is to bound the \emph{number of queries} performed in the above process. For this, we establish the following claim:

\begin{claim}\label{clm:queryBound}
    Across building the coarse estimator, the vertex flagging, and the $\log n$ rounds of error isolation, the total number of queries made is bounded by $O(n \log^3 n)$.
\end{claim}

\begin{proof}
    This follows by simple book-keeping. Building the coarse estimator requires $O(n \log^3 n)$ queries by construction. The flagging procedure similarly performs $O(n \log^3 n)$ queries.

    In each of the $\log n$ rounds of error isolation, we sample $O(ds \log n) = O(n \log^2 n)$ many random edges and query them. We also query all edges incident to the $s$ sampled right vertices, which requires at most $ds=O(n\log n)$ additional queries. Across all $\log n$ rounds, the error isolation procedure requires $O(n\log^3 n)$ queries,
\end{proof}

\subsection{Concluding the Algorithm}

Now we summarize the above and prove \cref{lem:phase}. We use the procedure described in this section. \cref{lem:repeatedIsolationPass} bounds the expected number of incorrect predictions. \cref{clm:edgeRemovalBound} bounds the number of removed edges, which are the postponed edges. \cref{clm:queryBound} provides the desired query bound. Finally, we can absorb the small failure probability into the expected number of incorrect predictions by observing that the number of incorrect predictions is always bounded by $n^2$. Similarly, if the proposed set of edge removals is $> f/2$, we can instead remove no edges, and again absorb this as a small failure probability (again, using the absolute bound of at most $n^2$ incorrect predictions).

\section{AI Disclosure}

ChatGPT 6-Astra was essential in proving the improved bounds. In particular, ChatGPT proved an $\tilde{O}(n^{4/3})$ query algorithm for generalized sorting which combined the prediction-repair algorithm of \cite{lu2020generalized} with a subsampling algorithm to estimate the rank of each element in the graph. ChatGPT was then unable to refine or improve this result, and it stagnated at $\tilde{O}(n^{4/3})$ queries. The authors subsequently identified a way to further reduce the number of queries by relying on careful implications in the graph structure, and, after several refinements, achieved $\tilde{O}(n^{6/5})$ many queries. 

ChatGPT 6-Astra Ultra/Pro and 6.1-Sol Ultra were then prompted dozens of times with these generalized sorting algorithms over the course of several weeks, tasked with finding an improvement. Several improvements were found, including first, an $\tilde{O}(n^{10/9})$ query algorithm, then an $\tilde{O}(n^{17/16})$, and then an $\tilde{O}(n^{19/18})$ query algorithm, before eventually discovering the $\tilde{O}(n)$ query algorithm presented here. In full disclosure, although these models were given the $\tilde{O}(n^{6/5})$ query algorithm developed by the authors, it is not clear whether the tools developed therein were essential for discovering the ultimate $\tilde{O}(n)$ query algorithm here. Indeed, the only shared pieces are merely high-level architectural choices like (a) doing efficient prediction-repair and (b) sampling vertices to estimate statistics of other vertices.

\bibliographystyle{alpha}
\bibliography{references}

\end{document}